## Title

# Inverse-designed photonic interfaces beyond eigenmode expansion limits

## Authors

Chengfei Shang,[1*†] Zhaoxi Chen,[1, 2†] Yifan Wu[1], Hanke Feng[1] and Cheng Wang[1*]

## Affiliations

[1]Department of Electrical Engineering & State Key Laboratory of Terahertz and Millimeter Waves, City University of Hong Kong, Kowloon, Hong Kong, China

[2]Rhinoptix Technology Limited. 1 Aicheng Street, Yuhang District, Hangzhou, China.

[†]*These authors contributed equally to this article*
[*]*chenshang2-c@my.cityu.edu.hk*
[*]*cwang257@cityu.edu.hk*

## Abstract

Photonic integrated circuits (PICs) enable optical systems with dramatically increased performance, cost-effectiveness, and scalability through enhanced light-matter interactions, high-density integration, and mass production. Due to the significant mode mismatch between various integrated photonic platforms and optical fibers, spot-size conversion interfaces with low-loss, compact footprint, and high manufacturability are essential. Conventional spot-size converters based on intuitive designs often require multi-layer tapering structures and tiny waveguide tips to adiabatically expand the eigenmodes. These rigid design constraints commonly lead to large device footprints and the requirements of multiple high-precision lithography steps. In this paper, we overcome these limitations using inverse design methods, which optimize the coupling efficiency over a large parameter space beyond traditional eigenmode evolution limits. Specifically, we demonstrate efficient and ultra-compact photonic interfaces on the thin-film lithium niobate (TFLN) platform, where the partially etched rib waveguides and non-vertical sidewalls have previously hindered the achievement of low-loss waveguide tapers in single-layer configurations. Our inverse-designed photonic structures achieve simulated and experimentally measured coupling efficiencies as low as 1 dB and 3 dB per facet between TFLN waveguides and lensed/ultra-high numerical aperture (UHNA) fiber, with broad 1-dB bandwidths exceeding 120 nm. The inverse-designed interfaces are highly compatible with standard TFLN PIC components and require only a single high-resolution lithography step. More importantly, the design concept transcends traditional eigenmode evolution theories and is broadly applicable to a variety of material platforms and application scenarios.

## Introduction

Photonic integration has emerged as a central theme in contemporary optics, with the potential not only to reduce the footprint and cost of optical systems, but also to dramatically enhance their performance and scalability. Analogous to electronic integrated circuits (ICs), these photonic integrated circuits (PICs) promise the mass production of numerous optical components on a single chip, thereby inspiring new and disruptive applications like photonic neural networks [1, 2], quantum information processors [3, 4], and co-packaged optical switches [5]. A prominent feature of PICs is the use of subwavelength-scale dielectric waveguides, and the resulting stronger electromagnetic confinement is one of core enablers of these innovations. However, the significantly reduced optical spot sizes in PICs also lead to substantial mode mismatches when interfacing them with standard optical fibers or between different photonic chips. A widely adopted solution is using waveguide tapers [6-9] based on the eigenmode-expansion (EME) theory, where the waveguides are adiabatically tapered until the eigenmode at the tip is matched with the intended output mode profile. Despite its effectiveness, this method typically requires long taper lengths (on the order of several hundred microns) to ensure a low-loss mode evolution, as well as precise and tiny tip sizes defined by high-resolution lithography processes. Furthermore, the optimal waveguide thickness for an ideal taper often differs from those of standard on-chip components, leading to the requirements of multi-layer tapering structures with further complication in device fabrication. Another popular fiber–chip coupling strategy is vertical grating coupler [10, 11], which converts guided modes into vertical radiation. However, due to the inherent vertical symmetry in popular photonic platforms, the coupling efficiency is usually limited to below 50% unless additional non-standard elements, such as bottom reflectors or double-layer designs [12, 13], are incorporated.

In emerging photonic platforms, for example the thin-film lithium niobate (TFLN) platform [14-16], the fiber–chip interface problem becomes even more formidable. By confining light in a sub-micron thick device layer, this platform not only inherits the excellent material properties of lithium niobate (LN), including low optical loss, large nonlinear susceptibility, and wide transparency window, but also features higher integration density and electro-optical (EO) modulation efficiency compared with its traditional counterpart. These advantages make TFLN PIC a promising solution for high-speed optical modulation [17-20], optical frequency comb generation [21-24], nonlinear frequency conversion [25-27], and microwave photonic systems [28-30]. However, typical TFLN waveguides adopt a half-etched rib structure with a total thickness of 500-600 nm for optimizing EO overlap, and a sidewall angle of 45–70° due to the nature of the physical dry-etching process. As a result, simply tapering the device layer only pushes the optical mode into the slab, and achieving eigenmodes with large mode field diameters (MFDs) generally requires multiple overlay processes to define two or more layers of tapers [31-33]. These delicate coupling structures demand lithography resolutions and alignment accuracies even beyond those required for functional TFLN devices. The lack of effective, low-cost, and fabrication-friendly interfaces has become a critical issue to TFLN PICs.

In recent years, inverse-design methods [34, 35] have demonstrated the ability to unlock nanophotonic designs that far exceed conventional limits imposed by human intuition. These approaches automatically explore large parameter spaces to optimize pre-defined objectives, enabling compact device footprints, enhanced robustness, and novel functionalities [36-38]. Recent efforts have further moved toward photonic system applications, such as particle accelerator [39], light detection and ranging [40], optical data transmitter [41], quantum information processing [42] and frequency comb source [43]. This methodology has also been successfully deployed in most common thin-film photonic platforms, including silicon [41], silicon nitride [44], silicon carbide [45], diamond [46], and TFLN [47]. However, previous demonstrations of inverse-designed interfaces [48, 49], have not yet gone beyond the traditional coupling theories, such as Bragg diffraction (for grating couplers) or eigenmode expansion (for waveguide tapers).

Here, we demonstrate a series of efficient inverse-designed on-chip interfaces that transcend traditional coupling theory boundaries. Unlike conventional approaches that rely on adiabatic EME, our topology-optimized couplers achieve a transient field evolution process that effectively expands the optical field profile in both vertical and horizontal dimensions. We apply this design strategy to tackle the coupling challenges in TFLN platform, achieving coupler designs that are fully compatible with standard TFLN components and require only a single high-resolution waveguide patterning process, while featuring simulated coupling loss as low as 1 dB/facet. We experimentally demonstrate low-loss interfacing of TFLN rib waveguides with lensed and ultra-high numerical aperture (UHNA) fibers, with measured coupling loss as low as 3 dB/facet and 1-dB operation bandwidths exceeding 120 nm. Our results present a brand-new photonic interfacing strategy that offers superior flexibility for interfacing various fiber types or photonic chip interconnections and are applicable to a wide range of photonic material platforms.

## Results

### Design and Simulation

Figure 1(a) presents a schematic comparison of our proposed inverse-designed photonic interface (bottom three) and the traditional bilayer inverse taper coupler (top) in the TFLN platform. In the traditional EME approach, the strongly confined waveguide mode is adiabatically transformed into a larger spot-size mode compatible with input/output fibers. Throughout this process, the optical field maintains eigenmode propagation, and the only way to enlarge the spot size in both vertical and horizontal directions is by forming waveguide tips that are small in both dimensions. For example, efficient coupling to a lensed fiber with a MFD of 2 μm requires a taper tip as small as 340 nm in width and 250 nm in height[31]. As a result, a double-layer taper strategy is generally required for TFLN rib waveguides. In these designs, the upper-layer rib is first tapered to push the optical mode into the bottom slab; subsequently, the bottom-layer waveguide is tapered to further expand the mode size. When the MFD further increases (e.g. 3.2 μm for UHNA fiber), the required tip width falls below 300 nm [32], and the fabrication process becomes increasingly complicated, often necessitating high-index cladding, triple-LN layers, or cladding waveguides. The multiple waveguide patterning steps and tiny taper tips put stringent demands on the lithography resolutions and alignment accuracies of the fabrication processes.

To address these limitations, we propose a different strategy: instead of sweeping cross-sectional taper geometries to match the eigenmode size, we treat the interface as a black box and employ inverse design to directly optimize the coupling efficiency through a transient field evolution process. A Gaussian beam of predetermined MFD is injected to the design region, which features the same rib height and slab thickness as the outgoing standard TFLN waveguide port. The pixelated elements of the design region can be freely chosen between rib and slab during the optimization process to maximize the objective function, i.e. the fundamental $TE_0$ mode power in the output rib waveguide. Figure 1(b) shows the 3D model of a representative optimized photonic interface. To visualize the field evolution process, we sliced the optical field distribution at six positions along the device, as shown in the insets. The results reveal that the beam area is significantly compressed in both vertical and horizontal directions through a transient field evolution. By reciprocity, light input from the waveguide side will be expanded and emitted into free space. The optimization was performed using the MEEP software package [50] with 3D finite-difference time-domain (FDTD) simulations and gradient-based topology optimization. All practical fabrication

constraints—including etching depth, crystal anisotropy, sidewall angle, and minimum feature size—were incorporated into the design process. Further mathematical details of the TFLN inverse-design methodology can be found in Supplementary Information Section S1.1 and related references [51-53].

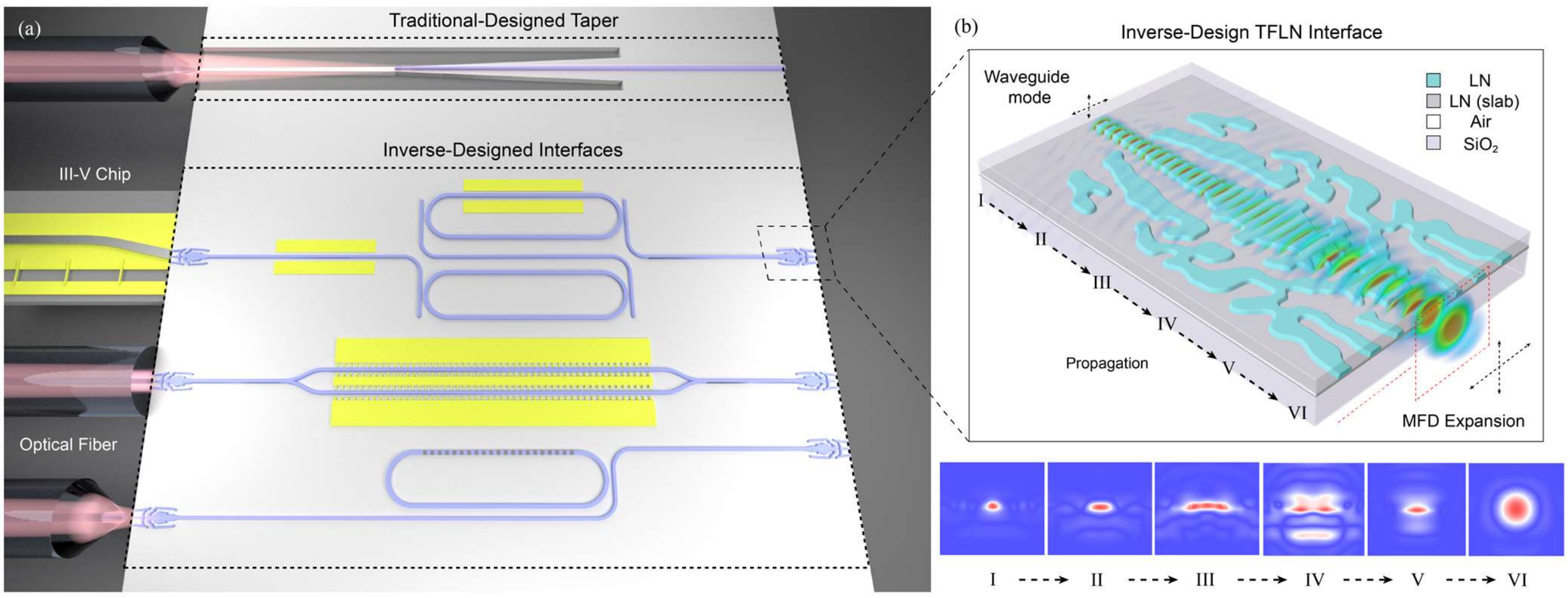

**Fig. 1.** Inverse-designed TFLN interface scheme. (a) Schematic illustrations of TFLN photonic circuits interfaced via traditional bi-layer (top) and the proposed inverse-designed couplers (bottom three). The inverse-designed interfaces feature footprints an order of magnitude shorter than the bilayer taper and are directly compatible with existing TFLN photonic components. (b) Zoom-in 3D view of a representative inverse-designed interface. The LN rib (cyan) and unetched slab (dark grey) are distinguished by different colors for clarity. Bottom insets show cross-sectional optical field evolution at six positions (I-VI) along the propagation direction, illustrating a clear MFD expansion process.

Based on this strategy, we first designed TFLN interfaces that can effectively couple light from lensed fiber (Figure 2(c)-2(e), MFD = 2 μm) and UHNA fiber (Figure 2(f)-2(h), MFD = 3.2 μm), into a 1-μm wide rib waveguide. The shared design parameters of these interfaces are shown in Figure 2(a). All devices were optimized by maximizing the average $TE_0$ coupling efficiency over a 100-nm wavelength window from 1500 to 1600 nm, evaluated at 1500, 1525, 1550, 1575 and 1600 nm. The devices were designed on a 500-nm x-cut TFLN layer with a 300-nm etching depth and a rib sidewall angle of 60° based on our typical TFLN fabrication process to be discussed later. During optimization, TFLN was modeled as an anisotropic dielectric with a refractive-index tensor [$n_o$, $n_e$, $\underline{n}_o$] along the $x$, $y$, and $z$ axes, where $n_o = 2.21$ and $n_e = 2.14$. The minimum feature size was set to 400 nm, constrained by the i-line stepper photolithography process. The footprint of each photonic interface is 25 × 15 μm², which compresses the propagation distance by more than an order of magnitude compared with conventional approaches while still providing sufficient design freedom for low-loss coupling (discussed in Supplementary Section S2.5). The silicon dioxide cladding ($h_{cld}$) and substrate thickness ($h_{box}$) were set to 1 and 2 μm, respectively. The silicon substrate directly beneath the devices is removed to suppress substrate leakage, improve the optimized coupling efficiency, and avoid obstruction from the diced silicon edge during fiber coupling; the corresponding analysis and discussion of the substrate removal is provided in Supplementary Section S2.4. The cladding thickness was optimized through a series of hyperparameter sweeps. Additional details of this hyperparameter analysis are provided in the Supplementary Information Section S2.1 to S2.3, which discusses key modeling considerations, including the influence of cladding thickness, substrate thickness, and beam size. We attribute the spot-size conversion to constructive interference of scattered light induced by algorithm-optimized dielectric structures, combined with total internal reflections at the air/$SiO_2$ boundaries of both upper and bottom claddings. The optimization was initialized from a uniform continuous design field (material density $\boldsymbol{\rho} = 0.5$) in the design region, providing an unbiased starting point between the rib and slab states. Figure 2(b) shows a typical iteration curve for one of our optimized photonic interfaces that contains three phases: continuous optimization for maximizing performance (210 rounds), geometric constraint that enforces the minimum feature size (300 rounds) and sidewall feature projection in the design region (80 rounds). Representative intermediate geometries from each optimization stage are provided in Supplementary Section S1.2. The total optimization commonly required 400-500 iterations in total, with each iteration taking approximately 4 minutes to complete on 32 CPU cores.

Figure 2(c)-2(h) presents the optimized interface patterns and their simulated performance over a 160-nm wavelength window (1480 nm-1640 nm) benchmarked against a 1-μm-wide bare waveguide. The minimum feature size was set to 400 nm to match our fabrication constraints. Both interfaces achieve a peak coupling efficiency of approximately –1 dB, corresponding to enhancement of 2 dB (lensed fiber) and 4.4 dB (UHNA fiber) relative to the bare waveguide cases. Considering typical chip insertion loss consists of contributions from at least two fiber-chip interfaces, these designs could in principle give rise to 2.5 – 7.6 times more fiber-to-fiber transmitted power. Across the optimization window, the efficiency variation remains within 1 dB. The performance degradation is slower at longer wavelengths, since dielectric confinement of the slab is weaker leading to better coupling. We note that the current TE-optimized design shows no significant coupling enhancement for the orthogonal TM mode (Supplementary Figure S11). However, polarization-insensitive operation with simultaneous TE/TM coupling enhancement can also be designed, an example of which is provided in Supplementary Information Section S4. Further improvements are also possible by reducing the minimum feature size to 200 nm, which would increase the peak coupling efficiency to –0.7 dB while remaining compatible with current photonic foundry capabilities (see design details in Supplementary Information Section S3). Overall, our designs demonstrate low coupling loss with an ultra-compact footprint, while significantly easing fabrication requirements and procedures. Device robustness is further analyzed in the Supplementary Information Section S5, including the effects of chip facet offsets, pattern erosion and dilation.

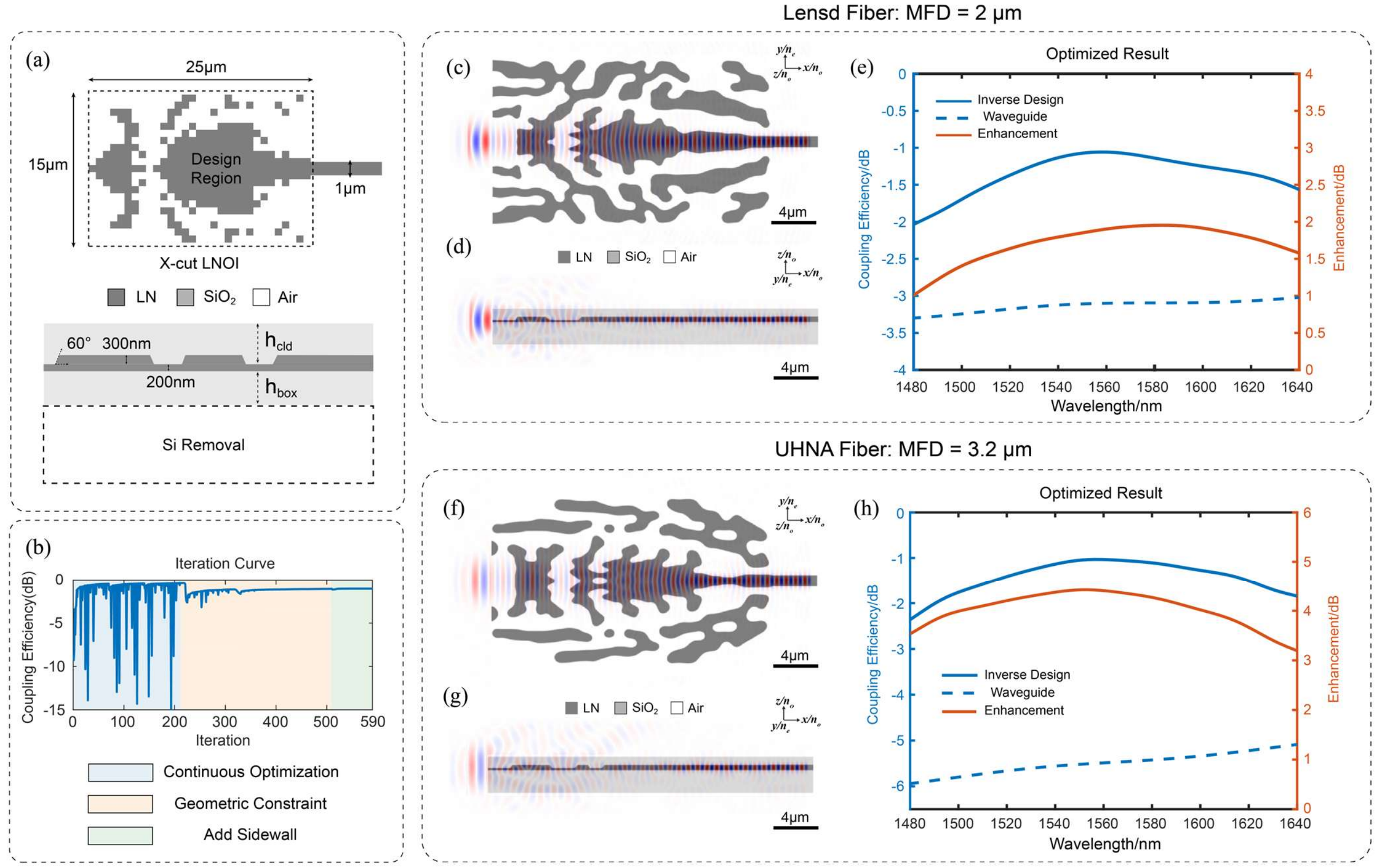

**Fig. 2.** Inverse-designed TFLN interfaces for coupling with lensed and UHNA fibers. (a) Shared design parameters of the interfaces, including the design region, LN thickness, etching depth, and sidewall angle. (b) A typical optimization curve of the photonic interface. Each optimization stage is indicated with different color. (c, f) Cross-sectional views (x-y plane) of the optimized design regions and simulated field distributions ($E_y$). (d, g) Cross-sectional views (x-z plane) and simulated field distributions ($E_y$). (e, h) Simulated coupling efficiencies of the designed interfaces (solid blue) and a bare waveguide (dashed blue), alongside the respective enhancement factors (red), showing high coupling efficiencies and broad operational bandwidths. Panels (c-e) and (f-h) correspond to simulation results for lensed and UHNA fibers, respectively.

## Fabrication & Characterization

We then experimentally validate the effectiveness of our designs by demonstrating TFLN photonic interfaces for lensed and UHNA fiber coupling. The corresponding fabrication flow is illustrated in Figure 3(a). All photonic devices were fabricated on commercial x-cut TFLN wafers (NANOLN) comprising the following structure: 500-nm single-crystal TFLN layer, 2-μm thermal oxide buffer layer, and 500-μm high-resistivity silicon substrate. The fabrication sequence starts with plasma-enhanced chemical vapor deposition (PECVD) of a 700-nm $SiO_2$ layer serving as an etch mask. Optical waveguide patterns and edge couplers (Figure 3(b)) were defined simultaneously using an ASML UV stepper lithography system (NFF facility, HKUST) followed by pattern transfer through sequential reactive ion etching (RIE) processes: first into the oxide mask layer, then into the LN layer with an etching depth of 300 nm. Residual etch mask was subsequently removed via wet etching. A 1-μm-thick PECVD $SiO_2$ upper cladding was deposited to complete the waveguide structure. To ensure a sharp chip facet right at the output boundary of our inverse-designed structures, we implemented a second, lower-resolution, lithography step. A thick photoresist mask (AZ2070) was patterned to shield the waveguide regions during subsequent deep etching stages, which sequentially remove the 1-μm $SiO_2$ cladding, 200-nm residual LN slab, and underlying 2-μm thermal oxide buffer layer in exposed areas through optimized RIE recipes. Following thorough solvent cleaning and de-ionized water rinse, the processed chips underwent precision dicing. Final device release was achieved through $XeF_2$ vapor-phase isotropic etching, selectively removing approximately 25 μm of silicon substrate beneath the edge couplers to create suspended structures compatible with fiber coupling (Figure 3(b), 3(c)).

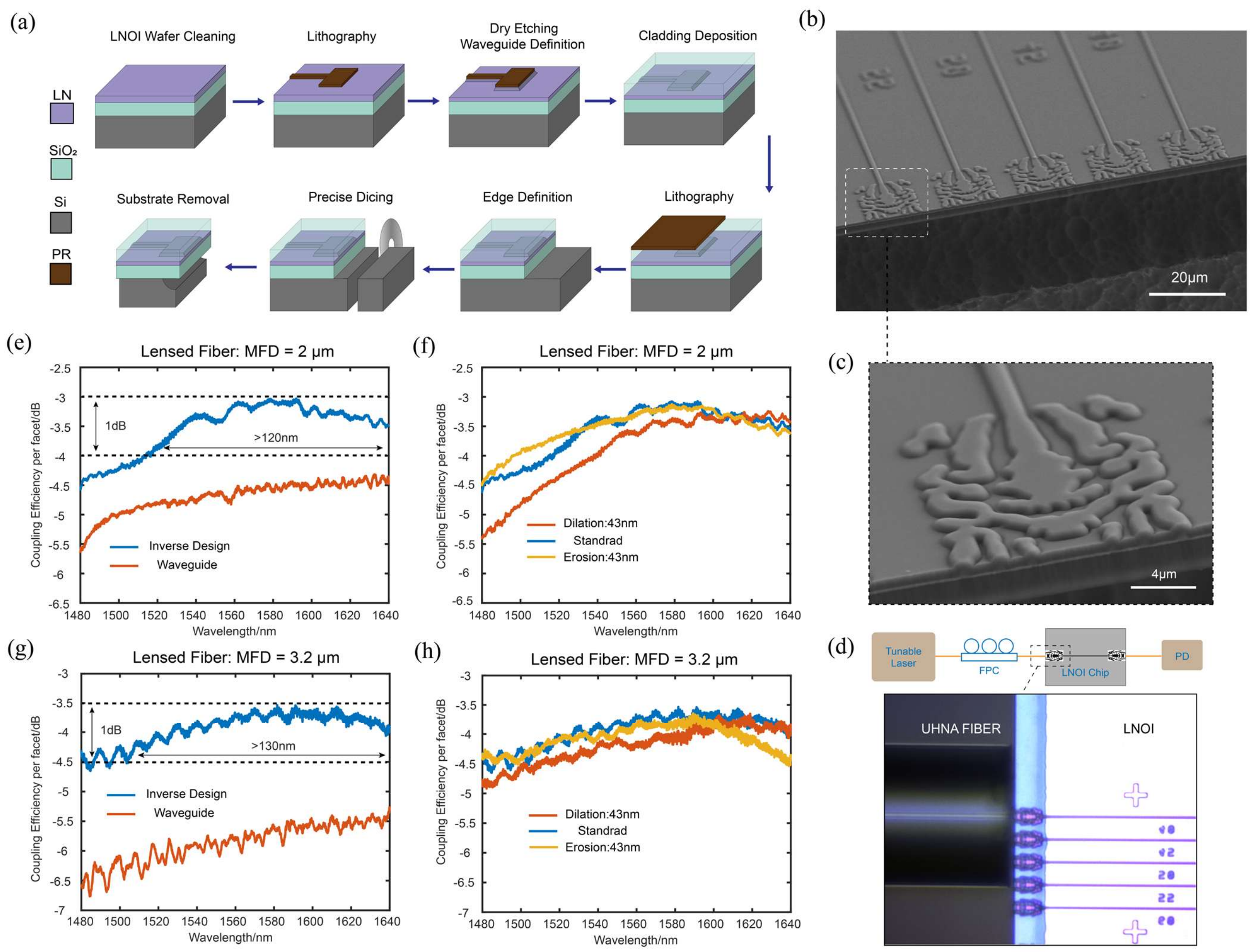

**Fig. 3.** Fabrication and characterizations of the inverse-designed photonic interfaces. (a) Schematic flowchart of the fabrication process. (b) Scanning electron microscope (SEM) image of an array of fabricated devices. (c) Magnified SEM view of a single device. (d) Experimental setup for measuring the coupling efficiency of photonic interface pairs; the inset camera image shows coupling between a UHNA fiber and a photonic interface. (e, g) Transmission spectra of inverse-designed interface pairs compared with a bare waveguide over the 1480–1640 nm wavelength range. The peak coupling efficiency reaches 3 dB/facet for lensed fiber (e) and 3.5 dB/facet for UHNA fiber (g), respectively. (f, h) Transmission spectra of photonic interfaces under fabrication variations, showing strong robustness against structural dilation/erosion of ±43 nm.

To avoid referencing uncertainties when measuring couplers only on one side, we characterize the devices by directly measuring the fiber-to-fiber transmission spectra with inverse-designed couplers at both input and output ports, using a characterization setup illustrated in Figure 3(d). Continuous-wave (CW) light from a tunable telecom laser (Santec TSL-710), with polarization control via a fiber polarization controller (FPC), was launched to the input coupler through a lensed or UHNA fiber. The output optical signal from a second coupler was collected by a second fiber and detected by a photodetector (PD, New Focus 1811). Figures 3(e) and 3(g) show the transmission spectra of the inverse-designed photonic interface pairs compared with a reference bare waveguide fabricated on the same chip. The coupling losses of the bare waveguide at 1580 nm were measured to be ~ –4.6 dB/facet and –5.7 dB/facet for lensed and UHNA fibers, respectively, consistent with previously reported values on the TFLN platform. In contrast, the coupling losses of the inverse-designed coupler pair were reduced to –3 dB/facet and –3.5 dB/facet at 1580 nm, representing fiber-to-fiber efficiency improvements of up to 4.4 dB compared with a bare waveguide case. The designed photonic interfaces also exhibit broadband performance, with a 1-dB bandwidth exceeding 120 nm (1520–1640 nm). To investigate the potential reason for the discrepancy between simulated and measured results, we have performed robustness analysis of our design in Supplementary Section S5.2. Specifically, for ±43 nm erosion/dilation, the simulated additional penalty remains below approximately 0.25 dB, although dilation produces a stronger red shift and larger degradation than erosion. Larger deviations, such as ±86 nm, lead to more noticeable penalties. These results suggest that our design is generally robust against fabrication variations but still benefit from more precise lithography and etching processes. We anticipate that employing fabrication processes with higher resolution, such as deep-UV stepper lithography or electron-beam lithography, could further enhance device performance. The remaining discrepancies between simulation and experiment are likely due to imperfections at small features, facet roughness, alignment errors, and oxide thickness non-uniformities.

**Discussions and Outlook**

To demonstrate the general applicability and transferability of our inverse-design approach, we further extended the methodology to several additional application scenarios. Figure 4 illustrates a series of TFLN photonic interfaces, optimized for coupling with commercial single-mode fiber (SMF) (Figure 4(a)-4(c)), III-V chips (Figure 4(d)-4(f)), and at visible wavelengths (MFD = 2 μm) (Figure 4(h)-4(j)).

Direct coupling with SMF remains highly challenging across nearly all integrated platforms due to the large MFD (~10 μm). To accommodate this, we adjusted our design parameters: the TFLN thickness was reduced to 300 nm and fully etched, the box and cladding thicknesses were increased to 4.7 μm, and the design region was enlarged to 50 μm × 20 μm. Although rib waveguide designs were abandoned due to the large spot-size conversion required, the simulated result reaches –1.3 dB/facet. To the best of our knowledge, this represents the most compact SMF edge-coupling interface proposed to date.

The hybrid integration of III–V gain media with TFLN enables a wide range of applications, including external cavity lasers (ECL) [(54)], mode-locked lasers (MLL) [(55)], and self-injection locking (SIL) [(56)]. For the III–V chips, we consider an input spot size of 4 μm × 2 μm, injected either normally or at an angle of 20° in air. The design parameters of these two III–V/TFLN interfaces are consistent with those presented in Figs. 2-3, i.e. standard TFLN rib structures. In the oblique-coupling case, the outgoing TFLN waveguide is tilted by 10°. The simulation results (peak coupling efficiency of –0.8/–1 dB) confirm the effectiveness of reducing coupling loss between the two chips within a compact footprint. We also observe that the oblique-coupling configuration outperforms the normally incident one, likely due to reduced facet reflections.

Nonlinear frequency generation, such as second-harmonic generation, represents another important application of TFLN photonics. Because of the shorter wavelength and stronger dielectric confinement, coupling visible light is significantly more challenging than in the telecom band. To address this, we designed a compact photonic interface (12 μm × 8 μm) on a 600-nm x-cut TFLN platform with a 250-nm etching depth, consistent with the parameters of common thin-film periodically poled lithium niobate (PPLN) devices that convert telecom light (1550 nm) to visible light (775 nm). The simulation results indicate a coupling efficiency of approximately –2.6 dB with a 2-μm MFD.

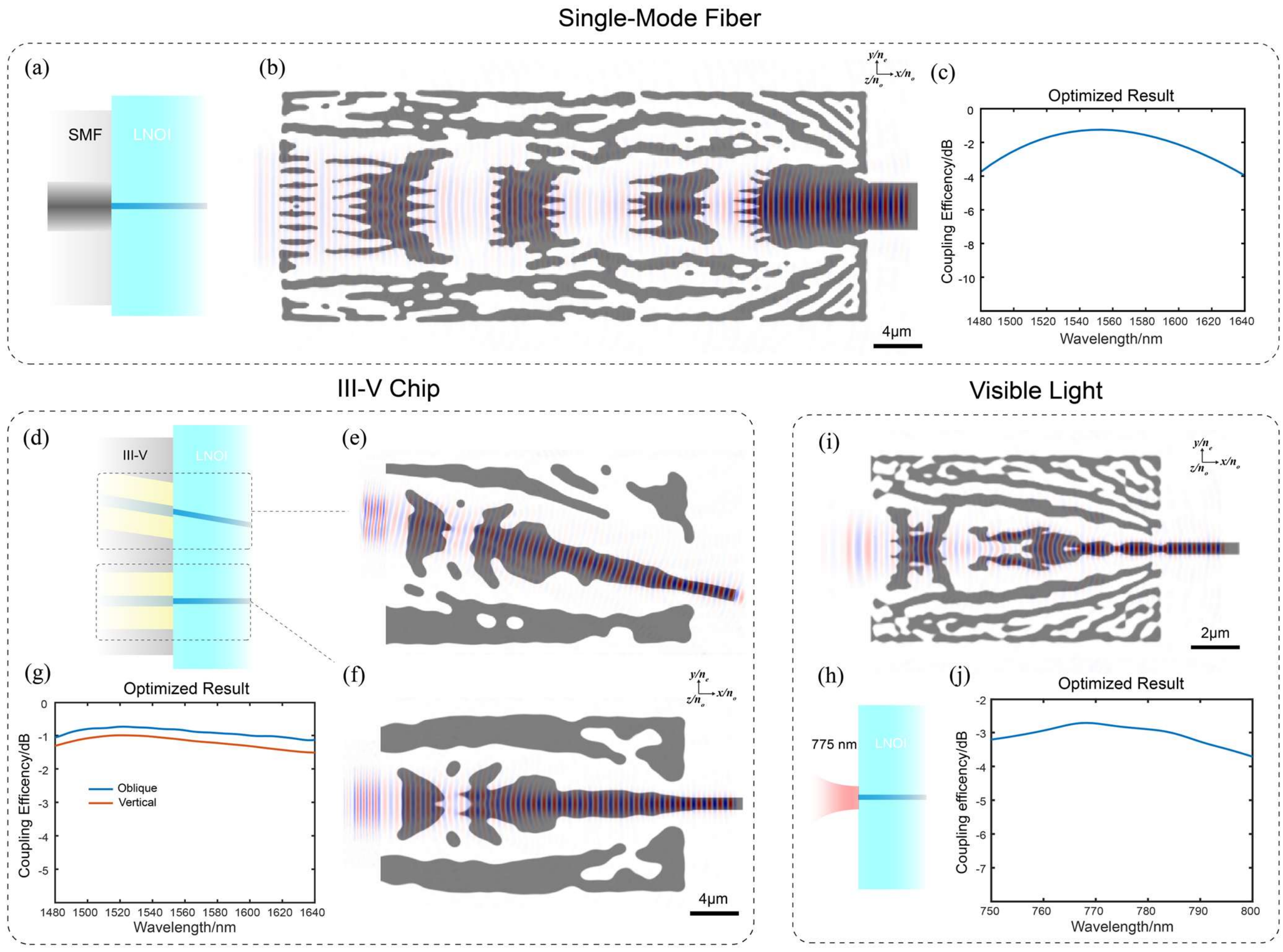

**Fig. 4.** Inverse-designed TFLN photonic interfaces for diverse application scenarios, including those coupling to (a-c) a single mode fiber, (d-g) normal- and oblique-incident III-V chips, and (h, j) operating at visible wavelengths. (a, d, h) Schematic illustrations of the respective scenarios. (b, e, f, i) Cross-section views (x-y plane) and their simulated field distributions ($H_z$). (c, g, j) Simulated coupling efficiencies of the corresponding designed interfaces.

In summary, we have experimentally demonstrated a new class of compact, low-loss, and fabrication-friendly photonic coupling interfaces enabled by inverse design. Compared with conventional waveguide tapers based on EME methods, our interfaces offer a significantly reduced footprint and a streamlined fabrication process, while eliminating the need for exhaustive parameter sweeps during design. The measured coupling efficiency in single-layer TFLN couplers reaches –3 dB per facet, accompanied by a 1-dB bandwidth of 120 nm. This device architecture and the associated design methodology hold promise for future applications in optical interconnects, telecommunications, and on-chip nonlinear optics. Furthermore, this approach is platform-agnostic and can be readily extended to other materials, including silicon, diamond, and silicon carbide.

During the revision of this manuscript, we note that similar inverse-designed approaches have been applied to photonic interfaces in the silicon nitride platform[57], where the focus is more on interfacing multiple waveguide eigenmodes with few-mode fibers.

## References

(1) Ashtiani, F.; Geers, A. J.; Aflatouni, F. An on-chip photonic deep neural network for image classification. *Nature* **2022**, *606* (7914), 501-506.
(2) Fu, T.; Zhang, J.; Sun, R.; Huang, Y.; Xu, W.; Yang, S.; Zhu, Z.; Chen, H. Optical neural networks: progress and challenges. *Light: Science & Applications* **2024**, *13* (1), 263.
(3) PsiQuantum, t. A manufacturable platform for photonic quantum computing. *Nature* **2025**, *641* (8064), 876-883.
(4) Aghaee Rad, H.; Ainsworth, T.; Alexander, R. N.; Altieri, B.; Askarani, M. F.; Baby, R.; Banchi, L.; Baragiola, B. Q.; Bourassa, J. E.; Chadwick, R. S.; Charania, I.; Chen, H.; Collins, M. J.; Contu, P.; D'Arcy, N.; Dauphinais, G.; De Prins, R.; Deschenes, D.; Di Luch, I.; Duque, S.; Edke, P.; Fayer, S. E.; Ferracin, S.; Ferretti, H.; Gefaell, J.; Glancy, S.; Gonzalez-Arciniegas, C.; Grainge, T.; Han, Z.; Hastrup, J.; Helt, L. G.; Hillmann, T.; Hundal, J.; Izumi, S.; Jaeken, T.; Jonas, M.; Kocsis, S.; Krasnokutska, I.; Larsen, M. V.; Laskowski, P.; Laudenbach, F.; Lavoie, J.; Li, M.; Lomonte, E.; Lopetegui, C. E.; Luey, B.; Lund, A. P.; Ma, C.; Madsen, L. S.; Mahler, D. H.; Mantilla Calderon, L.; Menotti, M.; Miatto, F. M.; Morrison, B.; Nadkarni, P. J.; Nakamura, T.; Neuhaus, L.; Niu, Z.; Noro, R.; Papirov, K.; Pesah, A.; Phillips, D. S.; Plick, W. N.; Rogalsky, T.; Rortais, F.; Sabines-Chesterking, J.; Safavi-Bayat, S.; Sazhaev, E.; Seymour, M.; Rezaei Shad, K.; Silverman, M.; Srinivasan, S. A.; Stephan, M.; Tang, Q. Y.; Tasker, J. F.; Teo, Y. S.; Then, R. B.; Tremblay, J. E.; Tzitrin, I.; Vaidya, V. D.; Vasmer, M.; Vernon, Z.; Villalobos, L.; Walshe, B. W.; Weil, R.; Xin, X.; Yan, X.; Yao, Y.; Zamani Abnili, M.; Zhang, Y. Scaling and networking a modular photonic quantum computer. *Nature* **2025**, *638* (8052), 912-919.
(5) Xiang, C.; Bowers, J. E. Building 3D integrated circuits with electronics and photonics. *Nature Electronics* **2024**, *7* (6), 422-424.
(6) Barwicz, T.; Peng, B.; Leidy, R.; Janta-Polczynski, A.; Houghton, T.; Khater, M.; Kamlapurkar, S.; Engelmann, S.; Fortier, P.; Boyer, N.; Green, W. M. J. Integrated Metamaterial Interfaces for Self-Aligned Fiber-to-Chip Coupling in Volume Manufacturing. *IEEE Journal of Selected Topics in Quantum Electronics* **2019**, *25* (3), 1-13.
(7) Mu, X.; Wu, S.; Cheng, L.; Fu, H. Y. Edge Couplers in Silicon Photonic Integrated Circuits: A Review. *Applied Sciences* **2020**, *10* (4), 1538.
(8) He, A.; Guo, X.; Wang, K.; Zhang, Y.; Su, Y. Low Loss, Large Bandwidth Fiber-Chip Edge Couplers Based on Silicon-on-Insulator Platform. *Journal of Lightwave Technology* **2020**, *38* (17), 4780-4786.
(9) He, A.; Guo, X.; Wang, T.; Su, Y. Ultracompact Fiber-to-Chip Metamaterial Edge Coupler. *ACS Photonics* **2021**, *8* (11), 3226-3233.
(10) Krasnokutska, I.; Chapman, R. J.; Tambasco, J. J.; Peruzzo, A. High coupling efficiency grating couplers on lithium niobate on insulator. *Optics Express* **2019**, *27* (13), 17681-17685.
(11) Cheng, L.; Mao, S.; Li, Z.; Han, Y.; Fu, H. Y. Grating Couplers on Silicon Photonics: Design Principles, Emerging Trends and Practical Issues. *Micromachines* **2020**, *11* (7), 666.
(12) Hammond, A. M.; Slaby, J. B.; Probst, M. J.; Ralph, S. E. Multi-layer inverse design of vertical grating couplers for high-density, commercial foundry interconnects. *Optics Express* **2022**, *30* (17), 31058-31072.
(13) Wang, H.; Zuo, Y.; Yin, X.; Chen, Z.; Zhang, Z.; Wang, F.; Hu, Y.; Zhang, X.; Peng, C. Ultralow-loss optical interconnect enabled by topological unidirectional guided resonance. *Science Advances* **2024**, *10* (12), eadn4372.
(14) Zhu, D.; Shao, L.; Yu, M.; Cheng, R.; Desiatov, B.; Xin, C. J.; Hu, Y.; Holzgrafe, J.; Ghosh, S.; Shams-Ansari, A.; Puma, E.; Sinclair, N.; Reimer, C.; Zhang, M.; Lončar, M. Integrated photonics on thin-film lithium niobate. *Advances in Optics and Photonics* **2021**, *13* (2), 242-352.
(15) Boes, A.; Chang, L.; Langrock, C.; Yu, M.; Zhang, M.; Lin, Q.; Loncar, M.; Fejer, M.; Bowers, J.;

Mitchell, A. Lithium niobate photonics: Unlocking the electromagnetic spectrum. *Science* **2023**, *379* (6627), eabj4396.
(16) Hu, Y.; Zhu, D.; Lu, S.; Zhu, X.; Song, Y.; Renaud, D.; Assumpcao, D.; Cheng, R.; Xin, C. J.; Yeh, M.; Warner, H.; Guo, X.; Shams-Ansari, A.; Barton, D.; Sinclair, N.; Loncar, M. Integrated electro-optics on thin-film lithium niobate. *Nature Reviews Physics* **2025**, *7* (5), 237-254.
(17) Wang, C.; Zhang, M.; Chen, X.; Bertrand, M.; Shams-Ansari, A.; Chandrasekhar, S.; Winzer, P.; Loncar, M. Integrated lithium niobate electro-optic modulators operating at CMOS-compatible voltages. *Nature* **2018**, *562* (7725), 101-104.
(18) Xu, M.; He, M.; Zhang, H.; Jian, J.; Pan, Y.; Liu, X.; Chen, L.; Meng, X.; Chen, H.; Li, Z.; Xiao, X.; Yu, S.; Yu, S.; Cai, X. High-performance coherent optical modulators based on thin-film lithium niobate platform. *Nature Communications* **2020**, *11* (1), 3911.
(19) Feng, H.; Zhang, K.; Sun, W.; Ren, Y.; Zhang, Y.; Zhang, W.; Wang, C. Ultra-high-linearity integrated lithium niobate electro-optic modulators. *Photonics Research* **2022**, *10* (10), 2366-2373.
(20) Chen, Y.; Feng, H.; Wang, Z.; Zhang, K.; Xie, X.; Zeng, Y.; Ren, Y.; Wang, C. Mono-drive single-sideband modulation via optical delay lines on thin-film lithium niobate. *Optica* **2025**, *12* (5), 666-673.
(21) Zhang, M.; Buscaino, B.; Wang, C.; Shams-Ansari, A.; Reimer, C.; Zhu, R.; Kahn, J. M.; Loncar, M. Broadband electro-optic frequency comb generation in a lithium niobate microring resonator. *Nature* **2019**, *568* (7752), 373-377.
(22) Wang, C.; Zhang, M.; Yu, M.; Zhu, R.; Hu, H.; Loncar, M. Monolithic lithium niobate photonic circuits for Kerr frequency comb generation and modulation. *Nature Communications* **2019**, *10* (1), 978.
(23) Zhang, K.; Sun, W.; Chen, Y.; Feng, H.; Zhang, Y.; Chen, Z.; Wang, C. A power-efficient integrated lithium niobate electro-optic comb generator. *Communications Physics* **2023**, *6* (1), 17.
(24) Chen, Z.; Zhang, Y.; Feng, H.; Zeng, Y.; Zhang, K.; Wang, C. Microwave-resonator-enabled broadband on-chip electro-optic frequency comb generation. *Photonics Research* **2025**, *13* (2), 426-432.
(25) Wang, C.; Langrock, C.; Marandi, A.; Jankowski, M.; Zhang, M.; Desiatov, B.; Fejer, M. M.; Lončar, M. Ultrahigh-efficiency wavelength conversion in nanophotonic periodically poled lithium niobate waveguides. *Optica* **2018**, *5* (11), 1438-1441.
(26) Lu, J.; Surya, J. B.; Liu, X.; Bruch, A. W.; Gong, Z.; Xu, Y.; Tang, H. X. Periodically poled thin-film lithium niobate microring resonators with a second-harmonic generation efficiency of 250,000%/W. *Optica* **2019**, *6* (12), 1455-1460.
(27) Li, X.; Li, H.; Wang, Z.; Chen, Z.; Ma, F.; Zhang, K.; Sun, W.; Wang, C. Advancing large-scale thin-film PPLN nonlinear photonics with segmented tunable micro-heaters. *Photonics Research* **2024**, *12* (8), 1703-1708.
(28) Feng, H.; Ge, T.; Guo, X.; Wang, B.; Zhang, Y.; Chen, Z.; Zhu, S.; Zhang, K.; Sun, W.; Huang, C.; Yuan, Y.; Wang, C. Integrated lithium niobate microwave photonic processing engine. *Nature* **2024**, *627* (8002), 80-87.
(29) Zhu, S.; Zhang, Y.; Feng, J.; Wang, Y.; Zhai, K.; Feng, H.; Pun, E. Y. B.; Zhu, N. H.; Wang, C. Integrated lithium niobate photonic millimetre-wave radar. *Nature Photonics* **2025**, *19* (2), 204-211.
(30) Tao, Z.; Wang, H.; Feng, H.; Guo, Y.; Shen, B.; Sun, D.; Tao, Y.; Han, C.; He, Y.; Bowers, J. E.; Shu, H.; Wang, C.; Wang, X. Ultrabroadband on-chip photonics for full-spectrum wireless communications. *Nature* **2025**, *645* (8079), 80-87.
(31) He, L.; Zhang, M.; Shams-Ansari, A.; Zhu, R.; Wang, C.; Marko, L. Low-loss fiber-to-chip interface for lithium niobate photonic integrated circuits. *Optics Letters* **2019**, *44* (9), 2314-2317.
(32) Hu, C.; Pan, A.; Li, T.; Wang, X.; Liu, Y.; Tao, S.; Zeng, C.; Xia, J. High-efficient coupler for thin-film lithium niobate waveguide devices. *Optics Express* **2021**, *29* (4), 5397-5406.
(33) Liu, X.; Gao, S.; Zhang, C.; Pan, Y.; Ma, R.; Zhang, X.; Liu, L.; Xie, Z.; Zhu, S.; Yu, S.; Cai, X. Ultra-

broadband and low-loss edge coupler for highly efficient second harmonic generation in thin-film lithium niobate. *Advanced Photonics Nexus* **2022**, *1* (01), 016001.
(34) Jensen, J. S.; Sigmund, O. Topology optimization for nano-photonics. *Laser & Photonics Reviews* **2010**, *5* (2), 308-321.
(35) Molesky, S.; Lin, Z.; Piggott, A. Y.; Jin, W.; Vucković, J.; Rodriguez, A. W. Inverse design in nanophotonics. *Nature Photonics* **2018**, *12* (11), 659-670.
(36) Piggott, A. Y.; Lu, J.; Lagoudakis, K. G.; Petykiewicz, J.; Babinec, T. M.; Vučković, J. Inverse design and demonstration of a compact and broadband on-chip wavelength demultiplexer. *Nature Photonics* **2015**, *9* (6), 374-377.
(37) Shen, B.; Wang, P.; Polson, R.; Menon, R. An integrated-nanophotonics polarization beamsplitter with 2.4 × 2.4 μm2 footprint. *Nature Photonics* **2015**, *9* (6), 378-382.
(38) Ahn, G. H.; Yang, K. Y.; Trivedi, R.; White, A. D.; Su, L.; Skarda, J.; Vučković, J. Photonic Inverse Design of On-Chip Microresonators. *ACS Photonics* **2022**, *9* (6), 1875-1881.
(39) Sapra, N. V.; Yang, K. Y.; Vercruysse, D.; Leedle, K. J.; Black, D. S.; England, R. J.; Su, L.; Trivedi, R.; Miao, Y.; Solgaard, O.; Byer, R. L.; Vučković, J. On-chip integrated laser-driven particle accelerator. *Science* **2020**, *367* (6473), 79-83.
(40) Yang, K. Y.; Skarda, J.; Cotrufo, M.; Dutt, A.; Ahn, G. H.; Sawaby, M.; Vercruysse, D.; Arbabian, A.; Fan, S.; Alù, A.; Vučković, J. Inverse-designed non-reciprocal pulse router for chip-based LiDAR. *Nature Photonics* **2020**, *14* (6), 369-374.
(41) Yang, K. Y.; Shirpurkar, C.; White, A. D.; Zang, J.; Chang, L.; Ashtiani, F.; Guidry, M. A.; Lukin, D. M.; Pericherla, S. V.; Yang, J.; Kwon, H.; Lu, J.; Ahn, G. H.; Van Gasse, K.; Jin, Y.; Yu, S. P.; Briles, T. C.; Stone, J. R.; Carlson, D. R.; Song, H.; Zou, K.; Zhou, H.; Pang, K.; Hao, H.; Trask, L.; Li, M.; Netherton, A.; Rechtman, L.; Stone, J. S.; Skarda, J. L.; Su, L.; Vercruysse, D.; MacLean, J. W.; Aghaeimeibodi, S.; Li, M. J.; Miller, D. A. B.; Marom, D. M.; Willner, A. E.; Bowers, J. E.; Papp, S. B.; Delfyett, P. J.; Aflatouni, F.; Vuckovic, J. Multi-dimensional data transmission using inverse-designed silicon photonics and microcombs. *Nature Communications* **2022**, *13* (1), 7862.
(42) He, L.; Liu, D.; Gao, J.; Zhang, W.; Zhang, H.; Feng, X.; Huang, Y.; Cui, K.; Liu, F.; Zhang, W.; Zhang, X. Super-compact universal quantum logic gates with inverse-designed elements. *Science Advances* **2023**, *9* (21), eadg6685.
(43) Lucas, E.; Yu, S.-P.; Briles, T. C.; Carlson, D. R.; Papp, S. B. Tailoring microcombs with inverse-designed, meta-dispersion microresonators. *Nature Photonics* **2023**, *17* (11), 943-950.
(44) Pita Ruiz, J. L.; Dalvand, N.; Menard, M. Integrated silicon nitride devices via inverse design. *Nature Communications* **2025**, *16* (1), 9307.
(45) Yang, J.; Guidry, M. A.; Lukin, D. M.; Yang, K.; Vuckovic, J. Inverse-designed silicon carbide quantum and nonlinear photonics. *Light: Science & Applications* **2023**, *12* (1), 201.
(46) Dory, C.; Vercruysse, D.; Yang, K. Y.; Sapra, N. V.; Rugar, A. E.; Sun, S.; Lukin, D. M.; Piggott, A. Y.; Zhang, J. L.; Radulaski, M.; Lagoudakis, K. G.; Su, L.; Vuckovic, J. Inverse-designed diamond photonics. *Nature Communications* **2019**, *10* (1), 3309.
(47) Shang, C.; Yang, J.; Hammond, A. M.; Chen, Z.; Chen, M.; Lin, Z.; Johnson, S. G.; Wang, C. Inverse-Designed Lithium Niobate Nanophotonics. *ACS Photonics* **2023**, *10* (4), 1019-1026.
(48) Sapra, N. V.; Vercruysse, D.; Su, L.; Yang, K. Y.; Skarda, J.; Piggott, A. Y.; Vuckovic, J. Inverse Design and Demonstration of Broadband Grating Couplers. *IEEE Journal of Selected Topics in Quantum Electronics* **2019**, *25* (3), 1-7.
(49) Melus, B. M.; Suelzer, J. S.; Reano, R. M. Experimental characterization and analysis of an adjoint method inverse design compact edge coupler solution. *Optics Express* **2025**, *33* (21), 43494-43504.
(50) Oskooi, A. F.; Roundy, D.; Ibanescu, M.; Bermel, P.; Joannopoulos, J. D.; Johnson, S. G. Meep: A

flexible free-software package for electromagnetic simulations by the FDTD method. *Computer Physics Communications* **2010**, *181* (3), 687-702.
(51) Hammond, A. M.; Oskooi, A.; Johnson, S. G.; Ralph, S. E. Photonic topology optimization with semiconductor-foundry design-rule constraints. *Optics Express* **2021**, *29* (15), 23916-23938.
(52) Pan, Y.; Christiansen, R. E.; Michon, J.; Hu, J.; Johnson, S. G. Topology optimization of surface-enhanced Raman scattering substrates. *Applied Physics Letters* **2021**, *119* (6), 061601.
(53) Zhou, M.; Lazarov, B. S.; Wang, F.; Sigmund, O. Minimum length scale in topology optimization by geometric constraints. *Computer Methods in Applied Mechanics and Engineering* **2015**, *293*, 266-282.
(54) Wang, S.; Lin, Z.; Wang, Q.; Zhang, X.; Ma, R.; Cai, X. High-Performance Integrated Laser Based on Thin-Film Lithium Niobate Photonics for Coherent Ranging. *Laser & Photonics Reviews* **2024**, *18* (10), 2400224.
(55) Guo, Q.; Gutierrez, B. K.; Sekine, R.; Gray, R. M.; Williams, J. A.; Ledezma, L.; Costa, L.; Roy, A.; Zhou, S.; Liu, M.; Marandi, A. Ultrafast mode-locked laser in nanophotonic lithium niobate. *Science* **2023**, *382* (6671), 708-713.
(56) Snigirev, V.; Riedhauser, A.; Lihachev, G.; Churaev, M.; Riemensberger, J.; Wang, R. N.; Siddharth, A.; Huang, G.; Mohl, C.; Popoff, Y.; Drechsler, U.; Caimi, D.; Honl, S.; Liu, J.; Seidler, P.; Kippenberg, T. J. Ultrafast tunable lasers using lithium niobate integrated photonics. *Nature* **2023**, *615* (7952), 411-417.
(57) Ruiz, J. L. P.; Gougeon, S.; Youssef, M.; Ung, B.; LaRochelle, S.; Ménard, M. Inverse-Designed Edge Couplers for Multimode Silicon Nitride Photonics. In *Optical Fiber Communication Conference*, 2026; Optica Publishing Group: p Tu2J. 5.

**Funding:**

Research Grants Council, University Grants Committee (CityU 11204523, STG3/E-704/23-N, C1002-22Y, CityU 11215024), Croucher Foundation (9509005)

**Acknowledgement:**

We thank Prof. Steven Johnson, Dr. TAO Yuansheng and Dr. ZHANG Ke for fruitful discussions and valuable help with the manuscript.

**Competing interests:**

Authors declare that they have no competing interests.